\documentclass[12pt]{iopart}

\usepackage[dvips]{epsfig}

\usepackage{amssymb}
\usepackage{amsfonts}

\begin{document}
\title{Some useful combinatorial formulae for bosonic operators}

\author{P Blasiak$^{a,b}$, K A Penson$^{a}$, A I Solomon$^{c}$, A Horzela$^b$,
 \\and  G H E Duchamp$^d$}
\address{\ \linebreak$^{a}$ Universit\'{e} Pierre et Marie Curie,\linebreak
Laboratoire   de  Physique   Th\'{e}orique  des  Liquides, CNRS UMR 7600\linebreak
Tour 24, $2^{i\grave{e}me}$ \'{e}tage, 4, place Jussieu, F 75252 Paris
Cedex 05, France\linebreak}
\address{$^{b}$ H. Niewodnicza{\'n}ski Institute of Nuclear Physics,
Polish Academy of Sciences\linebreak
ul. Radzikowskiego 152, PL 31-342 Krak{\'o}w, Poland\linebreak}
\address{$^{c}$ The Open University, Physics and Astronomy Department,\linebreak Milton Keynes MK7 6AA, United Kingdom\linebreak}
\address{$^{d}$ Institut Galil\'ee, LIPN, CNRS UMR 7030\linebreak
99 Av. J.-B. Clement, F-93430 Villetaneuse, France\linebreak}
\eads{\linebreak\mailto{blasiak@lptl.jussieu.fr},\mailto{ penson@lptl.jussieu.fr},\mailto{ a.i.solomon@open.ac.uk},\mailto{ andrzej.horzela@ifj.edu.pl}, \mailto{ gded@lipn-univ.paris13.fr} \linebreak}

\begin{abstract}
\vspace{1mm}\linebreak
We give a general expression for the normally ordered form of a function $F[\hat{w}(a,a^\dag)]$ where $\hat{w}$ is a function of
boson creation and annihilation operators satisfying $[a,a^\dag]=1$. The expectation value of this expression in a
coherent state becomes an  exact generating function of Feynman-type graphs associated with the zero-dimensional
Quantum Field Theory defined by $F(\hat{w})$. This enables one to enumerate explicitly the graphs of given order
in the realm of combinatorially defined sequences. We give several examples of the use of this technique, including the applications to Kerr-type and superfluidity-type hamiltonians.
\end{abstract}

\pacs{03.65, 05.30.Jp}

\maketitle

In the normally ordered form of a function $F(a,a^\dag)$ of boson creation and annihilation operators  all the
annihilation operators are moved to the right using the commutation relation $[a,a^\dag]=1$. The importance of the
normal form, denoted by $\mathcal{N}[F(a,a^\dag)]$ and satisfying $F(a,a^\dag)=\mathcal{N}[F(a,a^\dag)]$, is evident,
as with it the expectation values of $F(a,a^\dag)$ can be easily evaluated in such canonical  states as the vacuum
$|0\rangle$ and coherent states $|z\rangle=e^{-|z|^2/2}\sum_{n=0}^\infty z^n/\sqrt{n!}\ |n\rangle$, ($z$ complex
and $a^\dag a |n\rangle=n|n\rangle$). The role of the normal form in Quantum Field Theory (QFT) is pre-eminent
\cite{BjorkenDrell,Vasiliev}.

In this work we consider functions $F(\hat{w})$ that involve operators $\hat{w}(a,a^\dag)$ in the form of a
product of positive powers of $a^\dag$ and $a$, and powers of $(a+a^\dag)$, although the formulae derived below
are valid for more general $\hat{w}$'s. Our considerations are based on the following operational property of formal power series. Let $f(x)=\sum_{n=0}^\infty f_nx^n/n!$ and $g(x)=\sum_{n=0}^\infty g_nx^n/n!$ be two
formal power series, also called the exponential generating functions (egf) of sequences $\{f_n\}_{n=0}^\infty $
and $\{g_n\}_{n=0}^\infty $, respectively. Then it can be verified that
\begin{eqnarray}\nonumber
\left.f\left(\lambda\frac{d}{dx}\right)g(x)\right|_{x=0}=\left.g\left(\lambda\frac{d}{dx}\right)f(x)\right|_{x=0}=\sum_{n=0}^\infty
f_n\cdot g_n\ \frac{\lambda^n}{n!},
\end{eqnarray}
which we shall call the product formula (PF). This implies the following property satisfied by $F(\lambda
\hat{w})$ with indeterminate $\lambda$:
\begin{eqnarray}
F(\lambda \hat{w})=\left.F\left(\lambda\frac{d}{dx}\right)e^{x\hat{w}}\right|_{x=0},
\end{eqnarray}
which by taking the normal form of both sides becomes
\begin{eqnarray}\label{normalF}
\mathcal{N}[F(\lambda \hat{w})]=\left.F\left(\lambda\frac{d}{dx}\right)\mathcal{N}(e^{x\hat{w}})\right|_{x=0}.
\end{eqnarray}
Note that in Eq.(\ref{normalF}) a separation has been achieved between the functional aspect (defined by $F$) and
the operator aspect (defined by $\hat{w}$) of the normal ordering. Conventionally we implement
$\mathcal{N}(e^{x\hat{w}})$ by using the auxiliary symbol $:\ :$ with $\mathcal{N}(e^{x\hat{w}})\equiv \
:G_{\hat{w}}(x,a,a^\dag):$ , where under the symbol $:\ :$ the function $G_{\hat{w}}(x,a,a^\dag)$ is normally
ordered assuming that $a^\dag$ and $a$ {\em commute} \cite{KlauderSudarshan,Louisel}. Then
Eq.(\ref{normalF}) becomes
\begin{eqnarray}\label{normalF1}
\mathcal{N}[F(\lambda \hat{w})]=\left.F\left(\lambda\frac{d}{dx}\right):G_{\hat{w}}(x,a,a^\dag):\right|_{x=0}.
\end{eqnarray}
Note that the expression of Eq.(\ref{normalF1}) arises in the evaluation of the partition function $Z_\beta$ for the system defined by the hamiltonian $\mathcal{H}(\hat{w})$
\begin{eqnarray}\nonumber
Z_\beta={\textnormal{Tr}}\ e^{-\beta\mathcal{H}(\hat{w})}=\frac{1}{\pi}\int d^2 z\left[\left.e^{-\beta\mathcal{H}\left(\frac{d}{dx}\right)}G_{\hat{w}}(x,z,z^*)\right|_{x=0}\right],
\end{eqnarray}
taking the trace over the coherent state representation, $\beta=(k_BT)^{-1}$ \cite{Gilmore}.
The problem of finding $\mathcal{N}[F(\lambda\hat{w})]$ reduces to that of finding $\mathcal{N}(e^{x\hat{w}})$,
still however  a non-trivial task, {\em vide} the classical references \cite{KlauderSudarshan,Louisel,Wilcox}. We
have recently found expressions for $G_{\hat{w}}(x,a,a^\dag)$ for several types of operators $\hat{w}$
 of the form $\hat{w}_{(r,s)}=(a^\dag)^ra^s$ \cite{BPS} as well as for
$\hat{w}_{(\vec{r},\vec{s})}=\prod_{k=1}^M \hat{w}_{(r_k,s_k)}$ with $r,s,r_k,s_k$ positive integers \cite{MBPS}.   

At this point it is already possible to relate  Eq.(\ref{normalF1}) to enumerative formulae for Feynman-like
graphs in  QFT \cite{Bender}. Assume that our formal power series $F(x)$ can be written in the form
$F(x)=\exp\left( \sum_{m=1}^\infty L_m\frac{x^m}{m!}\right)$,
and we similarly assume that we may define operators $V_n^{(\hat{w})}(a,a^\dag)$ by
\begin{eqnarray}\label{V}
:G_{\hat{w}}(x,a,a^\dag):=\ :\exp\left(\sum_{n=1}^\infty V_n^{(\hat{w})}(a,a^\dag)\frac{x^n}{n!}\right):\ .
\end{eqnarray}
Explicit examples \cite{BPS} from which  the operators $V_n^{(\hat{w})}(a,a^\dag)$ may  be read off include:
\begin{equation}
\label{11}
\begin{array}{rcl}
\hat{w}=a^\dag a,&&\mathcal{N}[\exp\left(x a^\dag a\right)]=\ :\exp\left[a^\dag
a(e^x-1)\right]:\ ,
\end{array}
\end{equation}
\begin{equation}
\label{r1} 
\begin{array}{rcl}
\hat{w}=(a^\dag)^ra,&& \mathcal{N}[\exp\left(x (a^\dag)^ra\right)]=\\
&&=\
:\exp\left[a^\dag a\sum_{n=1}^\infty (a^\dag)^{(r-1)n}(r-1)^n\frac{\Gamma
(n+\frac{1}{r-1})}{\Gamma(\frac{1}{r-1})}\frac{x^n}{n!}\right]:
\end{array}
\end{equation}
with $r=2,3\ldots$ , as well as  more involved expressions for other $\hat{w}(a,a^\dag)$.
Thus, Eq.(\ref{normalF1}) may be written as

\begin{equation}
\label{normalLV}
\begin{array}{c}
\mathcal{N}[F(\lambda\hat{w})]
=\exp\left(\sum_{m=1}^\infty\frac{L_m}{m!}\lambda^m\frac{d^m}{dx^m}\right)\cdot\left.:\exp\left(\sum_{n=1}^\infty V_n^{(\hat{w})}(a,a^\dag)\frac{x^n}{n!}\right):\ \right|_{x=0}.
\end{array}
\end{equation}
We eliminate the  operators $a$ and $a^\dag$ by taking the matrix element of Eq.(\ref{normalLV}) in the coherent
state $|z\rangle$ and using $a|z\rangle = z|z\rangle$. This yields
\begin{equation}
\label{normalLVz}
\begin{array}{c}
\langle z|\mathcal{N}[F(\lambda\hat{w})]|z\rangle
=\left.\exp\left(\sum_{m=1}^\infty\frac{L_m}{m!}\lambda^m\frac{d^m}{dx^m}\right)\cdot\right.\left.
\exp\left(\sum_{n=1}^\infty V_n^{(\hat{w})}(z,z^*)\frac{x^n}{n!}\right)\right|_{x=0}.
\end{array}
\end{equation}
By specifying $z=1$ in Eq.(\ref{normalLVz}), defining $V_n^{(\hat{w})}(1,1)=V_n^{(\hat{w})}$,
$\mathbf{V}=\{V_n^{(\hat{w})}\}_{n=1}^\infty$ and $\mathbf{L}=\{L_m\}_{m=1}^\infty$ we obtain
\begin{eqnarray}\nonumber
&&Z(\mathbf{L},\mathbf{V},\lambda)\equiv \langle 1|\mathcal{N}[F(\lambda\hat{w})]|1\rangle\\&&\label{bender}
=\left.\exp\left(\sum_{m=1}^\infty\frac{L_m}{m!}\lambda^m\frac{d^m}{dx^m}\right)\right.\left.\cdot
\exp\left(\sum_{n=1}^\infty V_n^{(\hat{w})}\frac{x^n}{n!}\right)\right|_{x=0}
\end{eqnarray}
which is essentially the counting formula cited by Bender {\em et al.} \cite{Bender}.
Due to the symmetry of the PF ,   we have
$Z(\mathbf{L},\mathbf{V},\lambda)=Z(\mathbf{V},\mathbf{L},\lambda)$, which may facilitate the calculations.
Furthermore it can be demonstrated that for all the forms of $\hat{w}$ used here the sequence $\mathbf{V}$ consists of
positive integers.
The formula Eq.(\ref{bender}) was employed in \cite{Bender} as an enumerative tool for counting the Feynman-like
graphs in  zero-dimensional QFT models, where the values of all Feynman integrals are equal to one. Our derivation
sheds light on its quantum origin by tracing back its sources to the boson normal ordering problem. By specifying
the sets $\mathbf{L}$ and $\mathbf{V}$ one can attempt to produce a (in general divergent) power series expansion in
$\lambda$:
\begin{eqnarray}\label{Z}
Z(\mathbf{L},\mathbf{V},\lambda)=\sum_{n=0}^\infty A_n(\mathbf{L},\mathbf{V}) \frac{\lambda^n}{n!}\
\end{eqnarray}
in which $A_n(\mathbf{L},\mathbf{V})$ can be related to known objects.
To see that, recall the definition of the multivariate Bell polynomials $\mathbb{B}(\mathbf{h},u)$ related to a function
$h(x)=\sum_{n=1}^\infty h_n\frac{x^n}{n!}$ through ($\mathbf{h}=\{h_n\}_{n=1}^\infty$)
\begin{eqnarray}
e^{uh(x)}=\sum_{n=0}^\infty \frac{x^n}{n!}\sum_{k=1}^nu^k\mathbb{B}_{nk}(\mathbf{h})=\sum_{n=0}^\infty
\frac{x^n}{n!}\mathbb{B}_n(\mathbf{h},u),
\end{eqnarray}
where the coefficients of the expansion $\mathbb{B}_n(\mathbf{h},u)=\sum_{k=1}^nu^k\mathbb{B}_{nk}(\mathbf{h})$
depend only on $h_1,\ldots,h_n$. We refer to \cite{Comtet,Aldrovandi} for further properties of
$\mathbb{B}_n(\mathbf{h},u)$.

With $\mathbb{B}_n(\mathbf{f})=\mathbb{B}_n(\mathbf{f},1)$ we see that the coefficients $A_n=A_n(\mathbf{L},\mathbf{V})$ factorize
\begin{eqnarray}\label{A}
A_n=\mathbb{B}_n(\mathbf{L})\cdot \mathbb{B}_n(\mathbf{V})
\end{eqnarray}
which for given $\mathbf{L}$ and $\mathbf{V}$ can be worked out (see below).

The utility of Eqs.(\ref{bender}) and (\ref{A}) goes beyond the specific definition of initial $\hat{w}$, and this
is the philosophy of Ref.\cite{Bender} where it was suggested that $\mathbf{L}$ and $\mathbf{V}$ could
be treated as initial {\em input} for QFT models. From this perspective Eqs.(\ref{bender}) and (\ref{A})
provide the starting point for a Feynman-like graph representation of the coefficients $A_n$ in Eq.(\ref{Z}), where
$A_n$ counts the number of graphs with $n$ labelled lines. The graph construction rules are: a
line starts from a white dot, the {\em origin}, and ends at a black dot, the {\em vertex}. We further associate
strengths $V_k$ with each vertex receiving $k$ lines and multipliers $L_m$ with a white dot which is the origin of
$m$ lines. Counting  such graphs consists in
calculating their multiplicity due to the labelling of lines and the factors $L_m$ and $V_k$.

\begin{figure}
\vspace{1cm}
\begin{center}\resizebox{10.5cm}{!}{\includegraphics{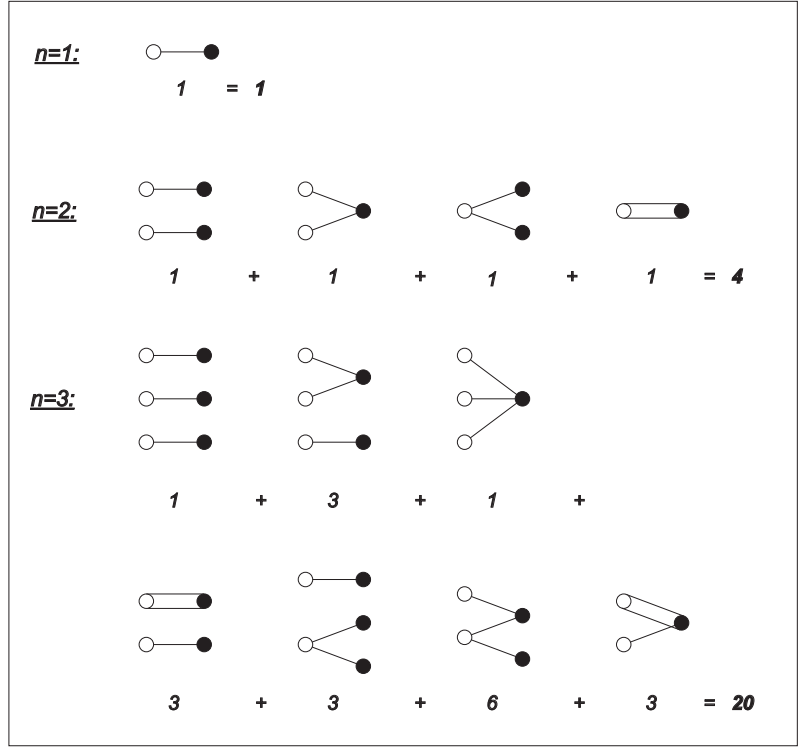}}
\caption{Lowest order Feynman-type graphs for Example 1 with $n=1,2,3$ lines. The number below each graph is its multiplicity.}\label{Fig1}
\end{center}
\end{figure}

We now specify  $\mathbf{L}$ and $\mathbf{V}$ and give some examples
of the explicit evaluation of $A_n$ along with the explicit graph representation:

\underline{Example 1:} $L_1=1$, $L_M=1$ ($M>1$), and $L_m=0$ otherwise, giving the function  $F(x)=\exp(x+x^M/M!)$; $V_n^{(\hat{w})}=1$
for $n=1,2\ldots$, which arises from the string  $\hat{w}=a^\dag a$, see Eq.(\ref{11}). This corresponds to  the
normal ordering problem  $\mathcal{N}[\exp(\lambda a^\dag a+\frac{\lambda^M}{M!}(a^\dag a)^M)]$. Note that the case $M=2$ describes the normal ordering of the exponential of the Kerr-type hamiltonian \cite{Knight} $\mathcal{H}=\lambda a^\dag a (1+\frac{\lambda}{2}a^\dag a)$. Using the
definition of the two variable Hermite-Kamp\'e de F\'eriet polynomials $H_n^{(M)}(x,y)$ (see \cite{Dattoli} and references therein)
\begin{eqnarray}
\sum_{n=0}^\infty H_n^{(M)}(x,y)\frac{t^n}{n!}=e^{xt+yt^M},
\end{eqnarray}
where $H_n^{(M)}(x,y)=n!\sum_{r=0}^{[n/M]}\frac{x^{n-Mr}y^r}{(n-Mr)!r!}$,
$F(x)$ can be expanded as
\begin{eqnarray}
F(x)=e^{x+\frac{x^M}{M!}}=\sum_{n=0}^\infty H_n^{(M)}(1,\frac{1}{M!})\frac{x^n}{n!}.
\end{eqnarray}
Eq.(\ref{Z}) yields
$A_n=H_n^{(M)}\left(1,\frac{1}{M!}\right)\cdot B_n$,
where the {\em Bell} numbers $B_n$ are defined through their egf: $\exp(e^x-1)=\sum_{n=0}^\infty B_n\frac{x^n}{n!}$
\cite{Bender,Comtet,Aldrovandi}. Observe that for $M=2$,
$H_n^{(2)}(1,\frac{1}{2})=\left(\frac{i}{\sqrt{2}}\right)^nH_n\left(-\frac{i}{\sqrt{2}}\right)=1,2,4,10,26,76,232,...$ are the {\em involution}
numbers \cite{Comtet} expressible using Hermite polynomials $H_n(x)$. The initial terms of
$A_n$ for $M=2$ are: $1,4,20,150,1352,15428,...$, see Fig.(\ref{Fig1}), and for $M=3$: $1,2,10,75,527,6293,...$, {\em etc}.
Note that whereas  $B_n$ counts all the partitions of an $n$-set, $H_n^{(M)}(1,\frac{1}{M!})$ counts partitions of
an $n$-set into  singletons and $M$-tons.

\underline{Example 2:} $L_m=m$ for $m=1,2\ldots$ , giving rise to $F(x)=\exp(\sum_{m=1}^\infty  {\scriptstyle m}
\frac{x^m}{m!})=\exp(xe^x)=\sum_{n=0}^\infty I_n\frac{x^n}{n!}$, where $ I_n=\sum_{k=0}^\infty \left(\!\!\!\!\!\stackrel{^{\ }_{\ }}{^{\ }}\right.\begin{array}{c}n\\k\end{array}\left.\!\!\!\!\!\stackrel{^{\ }_{\ }}{^{\ }}\right)k^{n-k}$
are {\em idempotent} numbers \cite{Comtet}. Again choosing   $V_n^{(\hat{w})}=1$, $n=1,2\ldots$, with
$\hat{w}=a^\dag a$, gives $A_n=I_n\cdot B_n=1,6,50,615,10192,214571\ldots$ \ . This corresponds
to normally ordering  $\mathcal{N}[\exp(\lambda (a^\dag a)e^{\lambda (a^\dag a)})]$.

\begin{figure}
\vspace{1cm}
\begin{center}\resizebox{11.5cm}{!}{\includegraphics{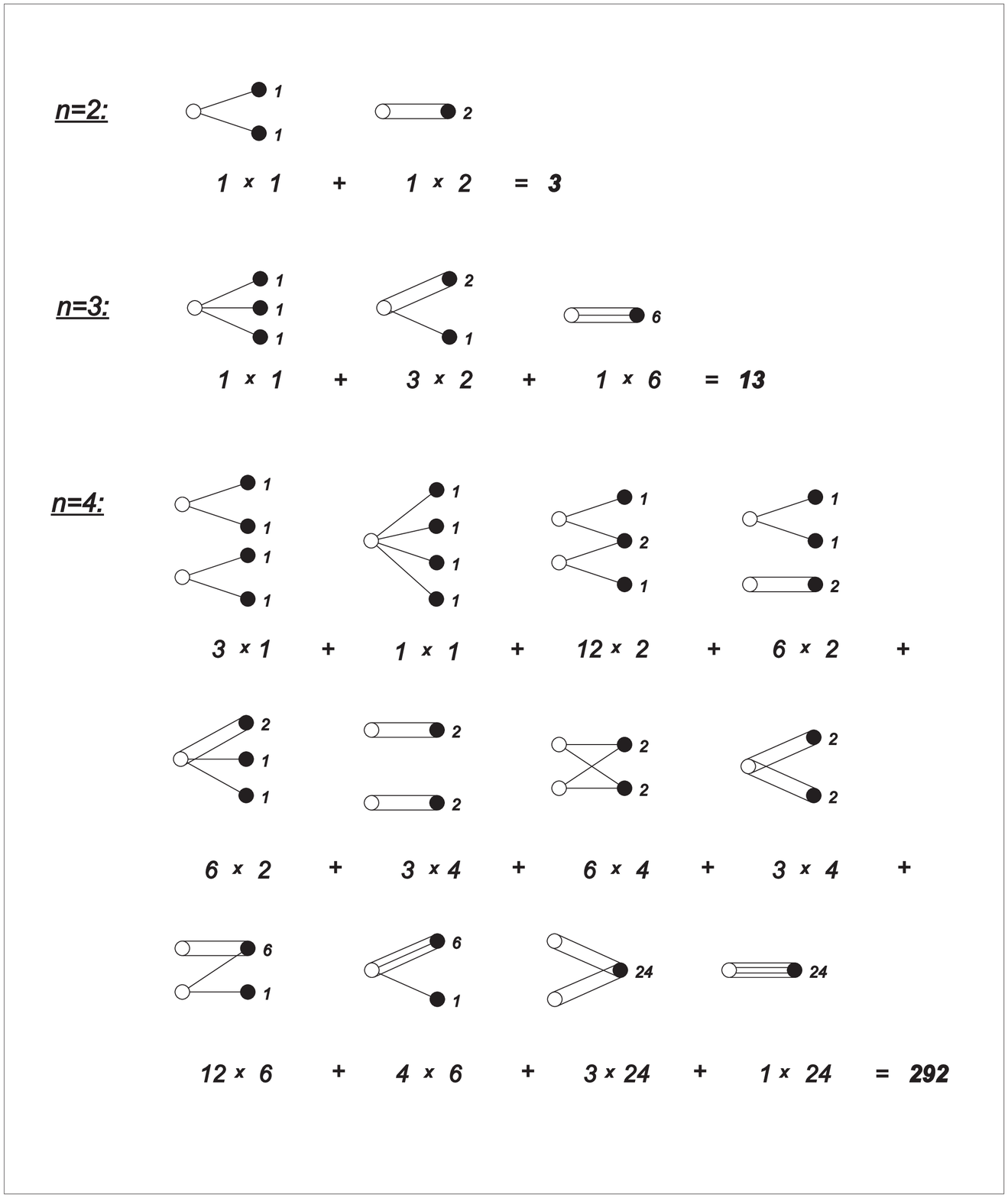}}
\caption{Lowest order Feynman-type graphs for Example 3 with $n=2,3,4$ lines. The number below each graph is $(multiplicity)\times\prod_k(vertex\ factor\ V_k=k!)$. Numbers at black dots (vertices) are vertex factors.}\label{Fig2}
\end{center}
\end{figure}

\underline{Example 3:} $L_1=0$, $L_m=1$ for $m=2,3\ldots$ , leading to  $F(x)=\exp(e^x-1-x)=\sum_{n=0}^\infty B_n^{(1)}\frac{x^n}{n!}$,
where $B_n^{(1)}$ are {\em restricted Bell} numbers which are defined as counting  partitions without singletons. (Note
that $B^{(1)}_n=\frac{1}{e}\sum_{k=0}^\infty \frac{(k-1)^n}{k!}$). Here we choose $V_n^{(\hat{w})}=n!$,
$n=1,2\ldots$ , derived from the string $\hat{w}=(a^\dag)^2a$, and producing via
$\exp\left(\frac{x}{1-x}\right)=\sum_{n=0}^\infty B_n^{(2,1)}\frac{x^n}{n!}$, (see Eq.(37) of \cite{BPS}),
$A_n=B_n^{(1)}\cdot B_n^{(2,1)}=0,3,13,292,5511,166091\ldots$, see Fig.(\ref{Fig2}). This corresponds to the normal ordering of
$\exp\left(e^{\lambda(a^\dag)^2a}-1-\lambda(a^\dag)^2a\right)$.

\underline{Example 4:} $L_{2m}=2m,\ L_{2m+1}=0$ for $m=0,1,2\ldots$ , giving $F(x)=\exp(x\sinh(x))$. If, as in Examples 1 and 2,
$V_n^{(\hat{w})}=1$, $n=1,2\ldots$, $\hat{w}=a^\dag a$, then by defining the {\em idempotent} polynomials
$I_n(t)=\sum_{k=0}^n\left(\!\!\!\!\!\stackrel{^{\ }_{\ }}{^{\ }}\right.\begin{array}{c}n\\k\end{array}\left.\!\!\!\!\!\stackrel{^{\ }_{\ }}{^{\ }}\right)k^{n-k}t^k$ we obtain $A_n=I_n^{(2)}\cdot
B_n=0,4,0,240,0,49938,0,24608160,0\ldots$ , where
$I_n^{(2)}=\sum_{k=0}^n\left(\!\!\!\!\!\stackrel{^{\ }_{\ }}{^{\ }}\right.\begin{array}{c}n\\k\end{array}\left.\!\!\!\!\!\stackrel{^{\ }_{\ }}{^{\ }}\right)(-1)^kI_k(-\frac{1}{2})I_{n-k}(\frac{1}{2})$, yielding
$\mathcal{N}[\exp(\lambda a^\dag a\sinh(\lambda a^\dag a))]$.

\begin{figure}
\vspace{1cm}
\begin{center}\resizebox{10.5cm}{!}{\includegraphics{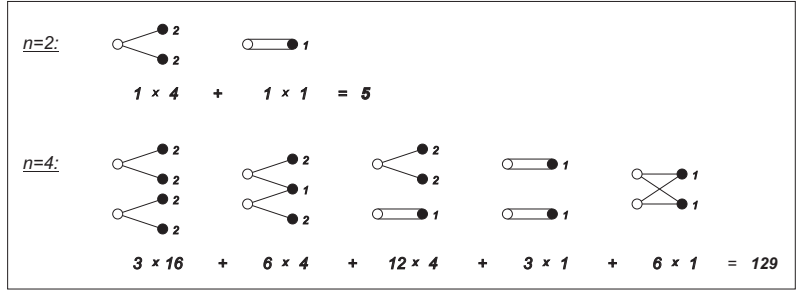}}
\caption{Lowest order Feynman-type graphs for Example~5 with $n=2,4$ lines. The number below each graph is $(multiplicity)\times\prod_k(vertex\ factor\  V_k=2\delta_{k,1}+\delta_{k,2})$.}\label{Fig3}
\end{center}
\end{figure}

\underline{Example 5:} In the last example we shall treat the function $\hat{w}=a+a^\dag$, using $F(x)=e^{x^M/M!}$, $M=1,2,3,...$. First observe that $\mathcal{N}(e^{x\hat{w}})=\ :G_{\hat{w}}(x,a,a^\dag):\ =\ :e^{x^2/2}e^{x(a+a^\dag)}:$ which is a consequence of the Heisenberg algebra. It follows that $V_1^{(\hat{w})}(a,a^\dag)=a+a^\dag$, $V_2^{(\hat{w})}(a,a^\dag)=1$ and $V_n^{(\hat{w})}(a,a^\dag)=0$ for $n>2$, see Eq.(\ref{V}), giving
$\mathbf{V}=\{2,1,0,0,...\}$ and $\mathbf{L}=\{\delta_{m,M}\}_{m=1}^\infty$. Let us define the modified Hermite polynomials $h_n(x)=\left(-\frac{i}{\sqrt{2}}\right)^nH_n\left(\frac{ix}{\sqrt{2}}\right)$ and then $\exp\left( 2x+x^2/2\right)=\sum_{n=0}^\infty \frac{h_n(2)}{n!}x^n$. Using Eqs.(\ref{Z}) and (\ref{A}) we get
\begin{eqnarray}\nonumber
Z_M(\mathbf{L},\mathbf{V},\lambda)
&=&\left.\exp\left(\frac{\lambda^M}{M!}\frac{d^M}{dx^M}\right)\cdot
\exp\left(2x+\frac{x^2}{2}\right)\right|_{x=0}\\
&=&\sum_{n=0}^\infty\frac{h_{Mn}(2)}{n!}\left(\frac{\lambda^M}{M!}\right)^n.\label{aa}
\end{eqnarray}
Starting with the simplest case $M\!=\!1$, the function $Z_1(\mathbf{L},\mathbf{V},\lambda)=\exp\left(2\lambda+\lambda^2/2\right)$ gives $A_n=h_n(2)=1,2,5,14,43,142,499,1850,...$, $n=0,1,2,...$ . The series of Eq.(\ref{aa}) can also be written down in closed form for $M=2$, corresponding to a single mode superfluidity-type hamiltonian $\mathcal{H}\sim(a+a^\dag)^2$ \cite{Solomon}, and for $M=3$ \cite{Gessel}:
\begin{eqnarray}\label{M2}
Z_2(\mathbf{L},\mathbf{V},\lambda)&=&\sum_{n=0}^\infty \frac{h_{2n}(2)}{n!}\left(\frac{\lambda^2}{2!}\right)^n=\frac{1}{(1-\lambda^2)^{1/2}}\exp\left(\frac{2\lambda^2}{1-\lambda^2}\right),\\
Z_3(\mathbf{L},\mathbf{V},\lambda)&=&\sum_{n=0}^\infty \frac{h_{3n}(2)}{n!}\left(\frac{\lambda^3}{3!}\right)^n\nonumber\\
&=&\frac{\exp\left(\displaystyle\phi^3\frac{\lambda^3}{6}-\phi^4\frac{\lambda^6}{8}\right)}{(1-\phi\lambda^3)^{1/2}}\ _2F_0\left(\frac{1}{6},\frac{5}{6};-;\frac{3\lambda^6}{2(1-\phi\lambda^3)^{3}}\right),
\end{eqnarray}
where $\phi(\lambda)=\frac{1-\sqrt{1-4\lambda^3}}{\lambda^3}$ and $ _2F_0$ is the generalized hypergeometric function (the Airy function, \cite{Abramowitz}). In these examples $Z_1$ and $Z_2$ are convergent series in $\lambda$ while $Z_3$ is not - summation of this series is understood in the generalized sense. From Eq.(\ref{aa}) we can read off the values of $A_n$: $A_{Mn}=\frac{(Mn)!}{(M!)^n n!}h_{Mn}(2)$ and zero otherwise, giving for $M=2$, $A_{2n}=1,5,129,7485,755265,116338005,...$, see Fig.(\ref{Fig3}).
Note, that whenever $Z$ is known in closed form the equation $Z(\lambda)=\left.\exp\left({\scriptstyle \lambda}\frac{d}{dx}\right)Z(x)\right|_{x=0}$ leads immediately to a set of graphs for which $L_m=\delta_{m,1}$. Thus for $M=2$, with Eq.(\ref{M2}) we have the following alternative descriptions: a) $L_m=\delta_{m,2}$; $V_1=2,\ V_2=1,\ V_{n>2}=0$ and b) $L_m=\delta_{m,1}$; $V_{2n}=(4n+1)(2n-1)!$ . However even if $Z$ is not known explicitly method a) leads to a simple, alternative, graphical description using Eq.(\ref{aa}).

In conclusion we see that the technique described herein and hinging on
Eq.(\ref{bender}) leads to a combinatorial and graphical description of many physical
systems.

\ack
We have benefited from the use of the EIS \cite{EIS} in the course of this work. One of us (PB) wishes to thank the Polish Ministry of Scientific Research and Information Technology for grant no: 1P03B 051 26.


\Bibliography{99}

\bibitem{BjorkenDrell} J.D.Bjorken and S.D.Drell, \emph{Relativistic Quantum Fields}, (McGraw and Hill, St. Louis, 1965).

\bibitem{Vasiliev} A.N.Vasiliev, \emph{Functional Methods in
Quantum Field Theory and Statistical Physics}, (Gordon and Breach, Amsterdam, 1998).

\bibitem{KlauderSudarshan} J.R.Klauder and E.C.G.Sudarshan, \emph{Fundamentals
of Quantum Optics}, (Benjamin, New York, 1968).

\bibitem{Louisel} W.H.Louisell, \emph{Quantum Statistical Properties of
Radiation}, (J.Wiley, New York, 1990).

\bibitem{Gilmore} W.M.Zhang, D.F.Feng and R.Gilmore, {\it Rev. Mod. Phys.} {\bf 62}, 867 (1990).

\bibitem{Wilcox} R.M.Wilcox, {\it J.Math. Phys.} {\bf 8}, 962 (1967).

\bibitem{BPS} P.Blasiak, K.A.Penson and A.I.Solomon,
{\it Phys. Lett. A} {\bf 309}, 198 (2003).

\bibitem{MBPS} M.A.M\'endez, P.Blasiak, K.A.Penson and A.I.Solomon, (unpublished).

\bibitem{Bender} C.M.Bender, D.C.Brody and B.K.Meister, {\it J.Math. Phys.} {\bf 40}, 3239 (1999);
C.M.Bender, D.C.Brody and B.K.Meister, {\it Twistor Newsletter} {\bf 45}, 36 (2000).

\bibitem{Comtet} L.Comtet, {\it Advanced Combinatorics} (Dordrecht,
Reidel, 1974).

\bibitem{Aldrovandi} R.Aldrovandi, \emph{Special Matrices of Mathematical
Physics}, (World Scientific, Singapore, 2001).

\bibitem{Knight} A.D.Wilson-Gordon, V.Bu\v{z}ek and P.L.Knight, {\it Phys.Rev. A} {\bf 44}, 7647 (1991).

\bibitem{Dattoli} G.Dattoli, S.Lorenzutta, C.Cesarano and
P.E.Ricci, {\it Integral Transform. and Spec. Funct.} {\bf 13}, 521 (2002).

\bibitem{Solomon} A.I.Solomon, {\it J.Math. Phys.} {\bf 12}, 390 (1971).

\bibitem{Gessel} I.M.Gessel and P.Jayawant, arXiv:math.CO/0403086 (2004).

\bibitem{Abramowitz} M. Abramowitz and I. A. Stegun, Eds., {\it Handbook of Mathematical Functions}, (Dover, new York, 1972), formula 10.4.59.

\bibitem{EIS} N.J.A.Sloane, {\it Encyclopedia of
Integer Sequences} (2004),
http://www.research.att.com/{\textasciitilde}njas/sequences.

\endbib
\end{document}